\documentclass[pre,eqsecnum,twocolumn, showpacs]{revtex4}

\usepackage{amsmath}
\usepackage{graphicx}
\usepackage{textcomp}

\newcommand{\rmd}{\mathrm{d}}

\newcommand{\kB}{k_\mathrm{B}}

\newcommand{\rTE}{r_\text{TE}}
\newcommand{\rTM}{r_\text{TM}}

\newcommand{\TE}{\mathrm{TE}}

\newcommand{\sigep}{4\pi\sigma}
\newcommand{\siSI}{\sigma^\mathrm{SI}}
\newcommand{\siSIe}{\siSI/\epsilon_0}
\newcommand{\ez}{\bar{\varepsilon}}
\newcommand{\ve}{\varepsilon}
\newcommand{\tG}{\tilde{\Gamma}}
\newcommand{\Am}{A_{\mu}}

\newcommand{\Li}{\mathrm{Li}}

\newcommand{\arsinh}{\mathrm{arsinh}}
\newcommand{\order}{\mathcal{O}}

\newcommand{\be}{\begin{equation}}
\newcommand{\ee}{\end{equation}}

\begin{document}

\title{Temperature correction to Casimir-Lifshitz free energy at low 
temperatures: semiconductors}
\date{\today}
\author{Simen A. \surname{Ellingsen}}\email{simen.a.ellingsen@ntnu.no}
\author{Iver \surname{Brevik}}\email{iver.h.brevik@ntnu.no}
\affiliation{Department of Energy and Process Engineering, Norwegian 
University of Science and Technology, N-7491 Trondheim, Norway}

\author{Johan S. H\o ye}\email{johan.hoye@phys.ntnu.no}
\affiliation{Department of Physics, Norwegian University of Science and 
Technology, N-7491 Trondheim, Norway}
\author{Kimball A. Milton}\email{milton@nhn.ou.edu}
\affiliation{Oklahoma Center for High Energy Physics and Department of Physics 
and Astronomy, The University of Oklahoma, Norman, OK 73019, USA}

\begin{abstract}
  The Casimir force and free energy at low temperatures have been the 
  subject of focus for some time. We calculate the temperature correction to 
  the Casimir-Lifshitz free energy between two parallel plates made of 
  dielectric material possessing a constant conductivity at low temperatures, 
  described through a Drude-type dielectric function. For the transverse 
  magnetic (TM) mode such a calculation is new. A further calculation for the 
  case of the TE mode is thereafter presented which extends and generalizes 
  previous work for metals. A numerical study is undertaken to verify the 
  correctness of the analytic results. 
\end{abstract}

\maketitle

There has been an explosion of interest in the Casimir effect \cite{casimir48},
generalized to dielectrics by Lifshitz \cite{lifshitz},
since the modern experiments began with Lamoreaux in 1997 \cite{lamoreaux}.
 The zero-temperature Casimir-Lifshitz theory seems to have been confirmed to 
 1\% accuracy over a range from 100 nm to a micrometer
 \cite{mohideen,roy,harris,ederth,chen,chan,bressi,bezerra,decca}.
 
 However, there has been a continuing controversy over the temperature dependence
 of this effect.  The prescription given in Ref.~\cite{schwinger} was seriously
 questioned by Bostr\"om and Sernelius \cite{bostrom} who pointed out that
necessarily the transverse electric reflection coefficient at zero frequency must vanish 
for metals. This discontinuity predicted a new linear temperature
term at low temperatures, resulting in about a 15\% correction to the result
found by Lamoreaux.  Lamoreaux believes that his experiment could not
be in error to this extent \cite{lamorpt}.  More heatedly, Mostepanenko
and collaborators have insisted that this behavior is inconsistent
with thermodynamics (the Nernst heat theorem), because it would predict,
for an ideal metal, that the free energy has a linear temperature
term at low temperature, and hence that the entropy would not
vanish at zero temperature \cite{klim01}.  Moreover, they assert
that the precision Purdue experiments rule out the linear temperature
term in the low temperature expansion \cite{decca}.

The issue is as yet unresolved, and is summarized in recent reviews
\cite{brevik06,klimchitskaya06a}.  We will not add further to the discussion
of this controversy here.  Rather, the purpose of this paper is to examine another 
purported temperature anomaly.  In several recent papers 
\cite{geyer05,klimchitskaya06,geyer06,klimchitskaya08} Geyer, Klimchitskaya, and Mostepanenko have 
claimed that in real dielectrics, which possess a very small, but nonzero conductivity 
which vanishes at $T=0$, a similar discontinuity in the transverse magnetic reflection 
coefficient occurs, which would lead to a similar violation of the Nernst theorem.
The same applies to semiconductors whose conductivity vanishes as temperature drops to 
zero. The solution according to these authors, as in the TE case for good conductors, is 
to prescribe the effect away.  We argue, however, that such a solution is physically 
unsatisfactory.

In Sec.~\ref{sec:2}, we will review and clarify their argument
for a standard Drude-type permittivity model for a weakly conducting material. 
We thereafter work out the leading-order temperature corrections to the free energy 
in the cases where the media are assumed to have a finite but small residual 
conductivity at $T=0$, as is implied when a Drude model is employed for taking the 
conductivity into account. This is a new result to our knowledge (a similar 
calculation for materials with zero conductivity was undertaken in Ref.~\cite{geyer05}). 
While this calculation does not solve the thermal anomaly brought forth in 
Refs.~\cite{geyer05,klimchitskaya06,geyer06} and reviewed in section \ref{sec:2}, 
it serves to further illuminate the mathematical behaviour of the free energy of 
poor conductors at very low temperatures when different models for the dielectric 
response of the materials are employed. A similar calculation is subsequently performed
for the TE mode, which extends that of Ref.~\cite{hoye07} in several ways: We 
allow for the conductivity to be small; we work out one further order of the temperature 
correction to the free energy; and we allow, for generality, the permittivity to 
have a finite dielectric constant term in addition to the Drude-type dielectric 
response due to free charges.

A word about units.  For our theoretical calculations, it is most convenient to
use Gaussian electromagnetic units, as well as natural space-time units:
$\hbar=c=\kB=1$.  However, for final results, which could be experimentally
observed, we use SI units.  The mapping between units is very simply carried
out by dimensional considerations, using the unit conversion factor
$\hbar c=1.97\times 10^{-5} \mbox{ eV\, cm}$.  The conductivity transformation
between Gaussian and SI units is the simple replacement 
$4\pi\sigma= \siSI/\epsilon_0$, where
$\epsilon_0=8.85\cdot 10^{-12}$ F/m is the absolute permittivity of
the vacuum and the notation $\siSI$ is used to explicate that SI units are used.

\section{Temperature Anomaly for Semiconductors}
\label{sec:2}
Here is a simple way to understand the argument of Ref.~\cite{geyer05}.
Suppose we model a dielectric with some small conductivity
by the permittivity function
\be\label{ep}
  \varepsilon(i\zeta)=1+\frac{\ez-1}{1+\zeta^2/\omega_0^2}+\frac{4\pi\sigma}{\zeta}.
\ee
 The essential point is that as $\zeta\to0$, $\varepsilon\to\ez$
if $\sigma=0$, otherwise $\varepsilon\to\infty$.
The Casimir (Lifshitz) free energy between two halfspaces, separated by a distance $a$,
assumed to be of the same material for simplicity, is given by
\begin{align}
F=\frac{T}{2\pi}{\sum_{m=0}^\infty}'\int_{\zeta_m}^\infty d\kappa\,\kappa
&[\ln(1-r_{\rm TM}^2e^{-2\kappa a})\notag\\
  &+\ln(1-r_{\rm TE}^2e^{-2\kappa a})].\label{F}
\end{align}
Here $\kappa,\rTE$, and $\rTM$ are functions of the discrete Matsubara frequencies 
$\zeta_m = 2\pi  m T$; $\kappa^2 = k_\perp^2+\zeta^2$ with $\mathbf{k}_\perp$ 
the transverse wave vector, directed parallel to the surfaces. As is conventional, 
the prime on the summation mark implies the $m=0$ term be taken with half weight. 
We need to examine the behavior of the reflection coefficients in the
small $\zeta$ limit.  These are
\begin{align}
r_{\rm TE}&=\frac{\kappa-\sqrt{\kappa^2+\zeta^2(\ve-1)}}
{\kappa+\sqrt{\kappa^2+\zeta^2(\ve-1)}},\label{rTE}\\
r_{\rm TM}&=\frac{\ve\kappa-\sqrt{\kappa^2+\zeta^2(\ve-1)}}
{\ve\kappa+\sqrt{\kappa^2+\zeta^2(\ve-1)}}
\end{align}
where $\ve=\ve(i\zeta)$. For the case of an ideal metal, it was $r^{\rm TE}$ which was 
discontinuous:
\be
  r_{\rm TE}(\zeta=0)=0, \quad \lim_{\zeta\to0}r_{\rm TE}= -1,
\ee
so this gave a linear temperature term when the sum over Matsubara frequencies
is converted to an integral according to the Euler-Maclaurin formula, for
example (Ref.~\cite{geyer05} uses the Abel-Plana formula, but that is 
equivalent).

For a dielectric the TE reflection coefficient is continuous and vanishes as $\zeta\to 0$,
but if there is a small (but not zero) conductivity which vanishes with $T$ linearly or 
faster, the TM coefficient exhibits a discontinuity at $\zeta=0$ as we now explain. 
When the conductivity is small we can assume there exists a temperature so that the 
$m=1$ Matsubara frequency, $\zeta_1=2\pi T$, satisfies the inequality
\be\label{discont}
  0<\sigep \ll \zeta_1 \ll \omega_0
\ee
in which case 
\be\label{finiteTdisc}
  r_{\rm TM}(i\zeta=0)=1,\quad r_{\rm TM}(i\zeta_1)=\frac{\ez-1}{\ez+1}.
\ee
Typical values of $\omega_0$ are in the optical or near IR frequency regions, so 
Eq.~(\ref{discont}) will hold at room temperature for many semiconductors. 
If now $\sigma$ goes to zero as $T\to 0$ linearly or faster, Eq.~(\ref{discont}) 
continues to hold true all the way to zero temperature where it becomes a true 
discontinuity:
\be\label{disc}
  r_{\rm TM}(i\zeta=0)=1,\quad \lim_{\zeta\to0}r_{\rm TM}(i\zeta)=\frac{\ez-1}{\ez+1}.
\ee
Clearly if $\sigma$ reaches some residual value $>0$, Eq.~(\ref{discont}) will not hold 
near zero temperature. Likewise the discontinuity disappears should $\sigma$ be exactly 
zero in a temperature region of finite width including $T=0$.

As in the metal case, Eq.~(\ref{disc}) gives rise to a linear temperature term in the
pressure and the free energy (see e.g.\ Ref.~\cite{brevik06} and references therein for 
details). Let $f_m$ be the summand of Eq.~(\ref{F}) or a similar expression for the 
Casimir pressure. Since $f_m$ is discontinuous at $m=0$, we must replace it by
a continuous function,
\be\label{fcont}
{\sum_{m=0}^\infty}'f_m=\frac12 f_0+\sum_{m=1}^\infty f_m=\frac12(f_0-\tilde f_0)
+{\sum_{m=0}^\infty}'\tilde f_m,
\ee
where $\tilde f_m$ is continuous, 
\be
\tilde f_0=\left\{ \begin{array}{rll} \tilde f_m &= \lim_{m\to0}f_m; & m=0 \\ 
\tilde f_m&=f_m, &m\neq 0 \end{array}\right.,
\ee 
so that the Euler-Maclaurin summation formula can be applied to the sum over 
$\tilde f_m$. Then the first term in the third form in Eq.~(\ref{fcont}) gives rise to a 
free-energy contribution which is a linear function of $T$. Defining the shorthand notation
\be\label{A0}
  A_0 = \left(\frac{\ez-1}{\ez+1}\right)^2,
\ee
that linear term is 
\begin{align}
  F^{\rm TM}&=\frac{T}{4\pi}\int_0^\infty d\kappa\,\kappa[\ln(1-A_0e^{-2\kappa a})
  -\ln(1-e^{-2\kappa a})]\nonumber\\
  &=\frac{ T}{4\pi}\sum_{n=1}^\infty \frac1{n}\left[A_0^n-1\right]
  \int_0^\infty d\kappa \,\kappa e^{-2n\kappa a}\nonumber\\
  &=\frac{ T}{16\pi a^2}\left[{\rm Li}_3(A_0)-\zeta(3)\right],
\end{align}
where the polylogarithmic function is
\be\label{polylog}
  {\rm Li}_n(\xi)=\sum_{k=1}^\infty \frac{\xi^k}{k^n}.
\ee
Note that the linear term vanishes for $\bar\varepsilon\to\infty$ as is clear from 
noting the relation to the Riemann zeta function:
\be
  \Li_n(1) = \zeta(n).
\ee
Thus at zero temperature, the entropy is nonzero,
\be
S=-\left(\frac{\partial F}{\partial T}\right)_V=-\frac1{16\pi a^2}
\left[\mbox{Li}_3(A_0)
-\zeta(3)\right],
\ee
which, if physical, is a violation of the Nernst heat theorem, or the third law
of thermodynamics, which states that the entropy of a system must vanish at
zero temperature.

\section{General formalism}\label{sec_general}

For reference throughout the next sections we will go though the formalism of determining 
the leading temperature corrections to the Casimir (Lifshitz) free energy by use of the 
Euler-Maclaurin formula, a procedure often employed previously.

Considering one polarization mode at a time, the free energy for the $q$ mode 
($q=$TM,TE) is written in the form
\be
  F_q = f(a,T){\sum_{m=0}^\infty}'g(m)
\ee
where we have pulled out a convenient prefactor.

When $T\to 0$ the Matsubara sum becomes an integral, so the temperature correction to the 
free energy, given by 
\be
  \Delta F_q = f(a,T)\left[{\sum_{m=0}^\infty}'-\int_0^\infty dm\right]g(m),
\ee
can be determined by use of the Euler-Maclaurin formula. For the summands of the 
Lifshitz formula, the higher derivatives of $g(m)$ are singular near $m=0$. When this 
is the case the Euler-Maclaurin formula can be applied to the sum starting at $m=1$ (or 
a higher value of $m$) instead, whereby
\begin{widetext}
  \begin{align}
\tG &\equiv\left[{\sum_{m=0}^\infty}'-\int_0^\infty dm\right]g(m) = \frac{1}{2}g(0)
-\int_0^1g(m)dm +\frac12g(1)- \sum_{k=1}^\infty \frac{B_{2k}}{(2k)!}g^{(2k-1)}(1)\notag \\
    &=\frac{1}{2}g(0)-\int_0^1g(m)dm+\frac{1}{2}g(1)-\frac{1}{12}g'(1)
    +\frac{1}{720}g'''(1)-\dots,\label{EM}
  \end{align}
\end{widetext}
where $B_n$ are the Bernoulli numbers,
\be
  B_2 = \frac1{6};~~B_4 = -\frac1{30}; ~~B_6 = \frac1{42};...
\ee
using the convention of \cite{abramowitz64} \S 23.2. [Two remarks are called for
here: We have assumed that $g$ and all its derivatives vanish at infinity, and
we have converted this formula into one which is commonly asymptotic because we have omitted
the remainder term which is present when only a finite number of derivatives
terms are retained.  Thus we are considering only the leading
terms in an asymptotic expansion for small $T$.]

As mentioned above, $g(m)$ is not analytic at $m=0$. It can be written in the asymptotic 
form for small $m$
\begin{align}
  g(m) \sim& c_0 + c_1 m + c_{3/2}m^{3/2}  \notag\\
  &+ c_{2l} m^2 \ln m + c_2 m^2 + \dots, \quad m\to 0.\label{ggeneral}
\end{align}

The terms needed for the right hand side of Eq.~(\ref{EM}) are now
\begin{subequations}
\begin{eqnarray}
  g(0)&=& c_0, \\
  g(1)&=& c_0 + c_1 + c_2+c_{3/2} + \dots, \\
  g'(1)&=& c_1  + c_{2l}+ 2c_2 + \frac{3}{2}c_{3/2} + \dots,\\
  g'''(1)&=& 2 c_{2l} - \frac{3}{8}c_{3/2} +  \dots,\\
  \int_0^1 dm g(m) &=& c_0+ \frac1{2}c_1 - \frac1{9}c_{2l}+ \frac1{3}c_2\nonumber\\
  &&\mbox{}+\frac{2}{5}c_{3/2} +\dots.
\end{eqnarray}
\end{subequations}
When inserted into Eq.~(\ref{EM}) the terms involving $c_0$ and $c_2$ cancel and 
one is left with
\begin{align}
  \tG \approx & -\frac{c_1}{12} + \frac{11c_{2l}}{360}
    -\frac{49}{1920}c_{3/2}+ \dots\label{generalCorr0}
\end{align}
Here the term due to $c_1$ is exact, whereas the terms referring to logarithms and 
half-integer powers of $m$, $c_{2l}$ and $c_{3/2}$, receive contributions from 
all higher derivatives in the Euler-Maclaurin formula, and to obtain exact expressions 
for the coefficients, all such terms must be kept, as we now show.

Retaining the higher derivative terms in the Euler-Maclaurin formula 
one finds by using
\begin{subequations}
\be
  \phi_{2n}=
\left.\frac{d^{2n-1}}{dm^{2n-1}}m^{3/2}\right|_{m=1} = -\frac{3 (4n-7)!}{2^{4n-5}(2n-4)!},
\quad n\geq 2,
\ee
and
\be
  \psi_{2n}=\left.\frac{d^{2n-1}}{dm^{2n-1}}m^2 \ln m\right|_{m=1}= 2 (2n-4)!,\quad 
  n\geq 2,
\ee
\end{subequations}
that the temperature correction to free energy is 
\be
  \Delta F_q = f(a,T)\tG,
\ee
where with  Eq.~(\ref{EM})
\be
  \tG =  -\frac{c_1}{12} + \Psi c_{2l}
    + \Phi c_{3/2}+  \dots,\label{generalCorr}
  \ee
with the coefficients 
\begin{subequations}
\begin{eqnarray}
  \Psi &=& \frac1{9}-\frac{B_2}{2}-\sum_{n=2}^\infty 
\frac{B_{2n}\psi_{2n}}{(2n)!},\label{Psi}\\
  \Phi &=& \frac1{2}-\frac{2}{5}-\frac{3B_2}{4}-\sum_{n=2}^\infty\
  \frac{B_{2n}\phi_{2n}}{(2n)!}.\label{Phi}
\end{eqnarray}
\end{subequations}
These series are formally divergent as is typical 
for perturbation series near singularities. Indeed, they arise from the asymptotic
Euler-Maclaurin formula (\ref{EM}).  For example, $\Phi$ can be recognized as a special
case of the expansion of Riemann zeta function in terms of Bernoulli numbers, Eq.~23.2.3 of
Ref.~\cite{abramowitz64} with an infinite number of terms retained in the sum, and
the remainder omitted.
A meaningful value can nonetheless be assigned to them through Borel summation as 
detailed in Appendix \ref{app}. Numerically, the mathematical software 
Maple computes the 
numerical values by means of a Levin u-transform to
\begin{subequations}
\begin{eqnarray}
  \Psi &= &0.03044845705840\dots\\
  \Phi &= &-0.0254852018898\dots
\end{eqnarray}
\end{subequations}
By either numerical or analytical correspondence we thus recognize that
\begin{subequations}
\begin{eqnarray}
  \Phi &=& \zeta(-\scriptstyle{\frac{3}{2}}),\label{anPhi}
\\
  \Psi &=& \frac{\zeta(3)}{4\pi^2},\label{anPsi}
\end{eqnarray}
\end{subequations}
where $\zeta$ is the Riemann zeta function.

When Eq.~(\ref{ep}) is used in the Lifshitz formalism with constant and finite $\sigma$ 
and $\ez>1$ in Secs.~\ref{sec_TMcond} and \ref{sec_TEcond}, we will find that the terms 
of $F$ stemming from $c_1, c_{3/2}$ and $c_{2l}$ are proportional to $T^2, T^{5/2}$ and 
$T^3$ respectively. Higher-order terms of $g(m)$ will likewise give higher-order 
temperature corrections.

\section{TM mode, residual conductivity}\label{sec_TMcond}

In the following sections we will work out the low temperature behaviour of 
corrections to 
the free energy under the assumption that a Drude-type dielectric function 
(\ref{ep}) may be used, and that $\sigma$ is finite and constant with respect to $\zeta$ 
and $T$ for small $T$ and $\zeta$. As argued in Ref.~\cite{ellingsen08}, 
when $\sigma$ is finite close to zero temperature  Nernst's theorem will be 
satisfied.  Here we will calculate explicitly the low temperature behaviour of the free 
energy for the TM mode. 

Conventionally, semiconductors are found within the broad interval of conductivity 
$\sigma$ in SI units $10^{-5} (\Omega \text{m})^{-1}<\siSI <10^5 (\Omega \text{m})^{-1}$,
that is
\be
  10^6 \text{s}^{-1} < \siSI/\epsilon_0 < 10^{16} \text{s}^{-1}.
\ee

For numerical purposes we will use the intermediate value 
$\sigma/\epsilon_0=10^{12}$s$^{-1}$, which is large enough not to hamper numerical 
verification unnecessarily, but small enough to distinguish the material in question 
from a good metal. The frequency corresponding to $\sigma/\epsilon_0$ for a metal is 
$\omega_p^2/\nu$, where $\omega_p$ is the plasma frequency and $\nu$ the relaxation 
frequency. For gold at room temperature $\omega_p^2/\nu$  has the approximate value 
$3.5\cdot 10^{22}$s$^{-1}$. 

Returning to Gaussian units, we consider the TM mode and introduce the shorthand notation
\begin{equation}\label{t}
  t=\frac{\zeta_1}{\sigep}=\frac{2\pi T}{\sigep} = \frac{2\pi \kB T}{\hbar (\siSIe)}
\end{equation}
and the symbol 
\begin{equation}
  \mu=mt. \label{mu}
\end{equation}
If $4\pi \sigma=10^{12}$ s$^{-1}$ as assumed above,
\be
  t\approx 0.83 T 
\ee
with $T$ in Kelvin. 

The free energy is given by Eq.~(\ref{F}), 
for which we now consider only the TM term:
\be\label{FTM}
  F^\text{TM}=\frac{T}{2\pi}{\sum_{m=0}^\infty}'\int_{\zeta}^\infty d\kappa\kappa 
  \ln(1-Ae^{ -2\kappa a}), 
\ee
where the reflection coefficient squared is
\begin{equation}
A\equiv r_\text{TM}^2=\left( \frac{\varepsilon-\sqrt{1+(\varepsilon-1)(\zeta/\kappa)^2}}
{\varepsilon+\sqrt{1+(\varepsilon-1)(\zeta/\kappa)^2}}\right)^2.
\end{equation}
Here and henceforth the index $m$ on Matsubara frequencies $\zeta_m$ and quantities 
dependent on it will frequently be suppressed.

The temperature corrections to 
the free energy at low temperatures are dominated by 
small frequencies, so we can assume as an approximation that the middle term of 
Eq.~(\ref{ep}) is simply equal to $\ez-1$ and write
\be\label{ep0}
  \ve(i\zeta) \approx \ez+\frac{\sigep}{\zeta} = \ez + \frac1{\mu}.
\ee
We define the dimensionless quantity
\begin{equation}\label{alpha}
  \alpha=2a(\sigep) = \frac{2a}{c}(\siSIe),
\end{equation}
where $a$ is the distance between the semiconductor plates.
For the value $\sigep\approx 10^{12}$s$^{-1}$ or smaller, $\alpha$ is a small quantity, 
which we use to define a criterion for the smallness of the conductivity in the remainder 
of this paper:
\be
  \alpha \ll 1.
\ee
For $a=1$\textmu m and $\sigma$ as above, as used for numerical purposes later, 
$\alpha$ has a value of about $6.7\cdot{10}^{-3}$, so this criterion is well satisfied.

By defining the variable $x$
\begin{equation}\label{x}
  x=2\kappa a = \frac{\kappa \alpha}{\sigep} = \frac{\kappa\alpha\mu}{\zeta}
\end{equation}
$A$ can be written
\be
  A = \left(\frac{1+\ez\mu-\mu\sqrt{1+[1+(\ez-1)\mu]\alpha^2\mu/x^2}}{1+\ez\mu
 +\mu\sqrt{1+[1+(\ez-1)\mu]\alpha^2\mu/x^2}}\right)^2,
\ee
and the integral (\ref{FTM}) with the use of Eq.~(\ref{t}) and $\zeta=2\pi m T$ becomes
\be
F^\text{TM}\equiv \frac{(\sigep)^3t}{4\pi^2\alpha^2}{\sum_{m=0}^\infty}'g(m), \label{xLif}
\ee
where
\begin{equation}
  g(m)=\int_{\alpha\mu}^\infty dxx\ln(1-Ae^{-x}). \label{gG}
\end{equation}

We wish now to extract explicitly the temperature dependence of the integrals $g(m)$ 
in Eq.~(\ref{gG}). The procedure we choose is to expand Eq.~(\ref{gG}) to leading order 
in the small parameter $\alpha$, and then expand the resulting term in powers of $m$ 
to get the form (\ref{ggeneral}). 

The first term in the Taylor expansion of the logarithm in powers of $\alpha$ is
\be\label{gNum}
  \ln(1-Ae^{-x}) = -\Li_1(\Am e^{-x}) + \order(\alpha^2)
\ee
where we use the polylogarithmic function defined in Eq.~(\ref{polylog}) and define the 
quantity
\be \label{Am}
  \Am = \left(\frac{1+(\ez-1)\mu}{1+(\ez+1)\mu}\right)^2.
\ee
For integral $s\leq 1$ the polylogarithm $\Li_s(y)$ can be expressed by elementary functions, specifically
\begin{align}
  \Li_1(y) &= -\ln(1-y); \quad\Li_0(y) = \frac{y}{1-y};\notag\\
  \Li_{-1} &= \frac{y}{(1-y)^2}.\label{Lielement}
\end{align}
The summand $g(m)$ thus has the form
\be
    g(m) = -\int_{\alpha\mu}^\infty dx\,x\Li_1(\Am e^{-x}) + \order(\alpha^2).\label{gI}\\
\ee

Now we will expand $g(m)$ in powers of $m$. It is easy to show from Eq.~(\ref{polylog}) 
that
\be\label{intpolylog}
  \int dy \Li_n(Ce^{-\beta y}) = -\frac1{\beta}\Li_{n+1}(Ce^{-\beta y});
\ee
from which by partial integration
\be
  g(m) = -\alpha\mu\Li_2(\Am e^{-\alpha\mu}) - \Li_3(\Am e^{-\alpha\mu})
+\order(\alpha^2).
\ee
We now use the property
\be\label{LiExpExp}
  \Li_n(Ce^{-y}) = \sum_{l=0}^\infty \frac{(-y)^l}{l!}\Li_{n-l}(C)
\ee
for $|C|<1$ to expand the polylogarithms in powers of $\alpha\mu$. 
The terms containing $\Li_2$ then cancel and we are left with
\be\label{gmprecise}
  g(m) = -\Li_3(\Am)+\frac{1}{2}\alpha^2\mu^2\Li_1(\Am)+\order(\alpha^2)
\ee
with $\Am$ given by Eq.~(\ref{Am}). Henceforth we shall denote the first two terms of the expansion (\ref{gmprecise}) $g_I(m)$ and $g_{II}(m)$. The remaining $\order(\alpha^2)$ term comes
from the error in Eq.~(\ref{gNum}).
As before we are going to truncate the expansion in $\alpha$ at leading order, 
but will evaluate the explicit correction $\sim\alpha^2$ to (\ref{gmprecise}) later as a 
measure of the error. We thus have the simple expression
\be\label{gm}
  g_I(m) = -\Li_3 (\Am).
\ee

We will next expand Eq.~(\ref{gm}) in $\mu$. $\Li_3(\Am)$ does not have a Taylor expansion 
near $m=0$ (where $A_0=1$) because its second derivative is singular here. Using 
\be
  \frac{\rmd}{\rmd y}\Li_n (y) = \frac{1}{y}\Li_{n-1}(y),
\ee
we differentiate Eq.~(\ref{gm}) to find
\be
  g_I'(m) = \frac{4t\Li_2 (\Am)}{[1+(\ez+1)\mu][1+(\ez-1)\mu]}.
\ee
We can use the identity \cite{lewin91}
\be
  \Li_2(z)+\Li_2(1-z) = \frac{\pi^2}{6} - \ln (z) \ln(1-z),
\ee
which is easily verified by differentiation, use of Eq.~(\ref{polylog}) and $\Li_2(1)=\pi^2/6$. 
Furthermore $\Li_2 (1-\Am)$ has a simple Taylor expansion around $\Am=1$,
\be
  \Li_2(1-\Am) = 4\mu - 4(\ez+1)\mu^2+\dots
\ee
and
\be
  \frac{4}{[1+(\ez+1)\mu][1+(\ez-1)\mu]} = 4-8\ez \mu+\dots,
\ee
whereby we find
\be\label{g'm}
  g_I'(m) =  \frac{2\pi^2t}{3} - 4m t^2\left(\frac{\ez\pi^2}{3}+4\right) + 
  16m t^2\ln 4\mu+\dots,
\ee
where the next term of the series is of order $t^3$.

Comparing with Eq.~(\ref{ggeneral}) we recognize the coefficients
\be
  c_1 = \frac{2\pi^2 t}{3},\quad c_{2l} = 8t^2,
\ee
which we insert into Eq.~(\ref{generalCorr}) to find
\be
  \left[{\sum_{m=0}^\infty}'-\int_0^\infty dm\right]g_I(m) = 
  -\frac{\pi^2t}{18} + 8\Psi t^2.
\ee

We thus obtain the approximate correction to the free energy for small $t$:
\begin{align}\label{dFTM}
  \Delta F_I^\text{TM} &= \frac{(\sigep)^3t}{4\pi^2\alpha^2}
  \left[{\sum_{m=0}^\infty}'-\int_0^\infty dm\right]g_I(m)\notag\\
  &\approx -\frac{(\sigep)^3}{72\pi^2 \alpha^2} t^2 \left[\pi^2 - 144\Psi t\right]
\end{align}
in terms of our reduced units $t$ and $\alpha$. In SI units inserting (\ref{anPsi}):
\begin{align}
  \Delta F_I^\text{TM} &= -\frac{\pi^2(\kB T)^2}{72\hbar (\siSIe)a^2} 
  + \frac{\zeta(3)(\kB T)^3}{\pi[\hbar (\siSIe)a]^2} \notag\\
  &= -\frac{\pi^2(\kB T)^2}{72\hbar (\siSIe)a^2}
  \left(1 -\frac{72\zeta(3)\kB T}{\pi^3\hbar \siSIe}\right).\label{dFTMSI}
\end{align}

\subsection{Correction due to subleading terms of Eq.~(\ref{gmprecise})}

Twice in the above we truncated the expressions at leading order in the parameter 
$\alpha$, in Eq.~(\ref{gNum}) and Eq.~(\ref{gmprecise}). As an indication of the 
magnitude of the error we will calculate the next order in $\alpha$ of 
Eq.~(\ref{gmprecise}) while a similar calculation for Eq.~(\ref{gNum}) 
is more troublesome due to singularities and 
beyond the scope of the present effort. The correction $\propto \alpha^2$ of 
Eq.~(\ref{gmprecise}) was
\be
  \Delta g(m) = \frac{1}{2}\alpha^2m^2t^2\Li_1(\Am)+\order(\alpha^3t^3).
\ee
We will only consider the first term, since the next terms give temperature corrections 
$\propto T^4$ and higher. We Taylor expand as before in powers of $\mu$
\be\label{Li1rel}
  \Li_1(\Am) = -\ln(4t)-\ln m+(\ez+2)\mu+\order(\mu^2),
\ee
wherewith the leading correction from $\Delta g_I(m)$ is found from Eq.~(\ref{generalCorr})
to order $T^3$ to which only the term $\propto m^2\ln m$ contributes:
\begin{align}
  \delta F^\text{TM} &= \frac{(\sigep)^3t}{4\pi^2\alpha^2}\delta\tG
  \approx \frac{(\sigep)^3\Psi t^3}{8\pi^2} .\label{lolimCorr}
\end{align}
Being $\alpha$ independent, the correction (\ref{lolimCorr}) is much smaller than the 
leading term (\ref{dFTM}) for small $\alpha$. In SI units:
\be\label{totalCorr}
  \delta F^\text{TM} = \frac{\zeta(3)(\kB T)^3}{4\pi\hbar^2c^2}+\order(T^4)
\ee
The relative magnitude of this term compared to the $T^3$ term of Eq.~(\ref{dFTMSI}) is 
with our numerical data
\be
  \frac{(\siSIe)^2 a^2}{4c^2}\approx 2.8\cdot 10^{-6}.
\ee
The correction from the truncation of Eq.~(\ref{gNum}) is likely to be of similar 
size and therefore much smaller than the accuracy of the numerical investigation.

\subsection{Numerical investigation of TM mode result}

\begin{figure}[tb]
  \begin{center}
    \includegraphics[width=3in]{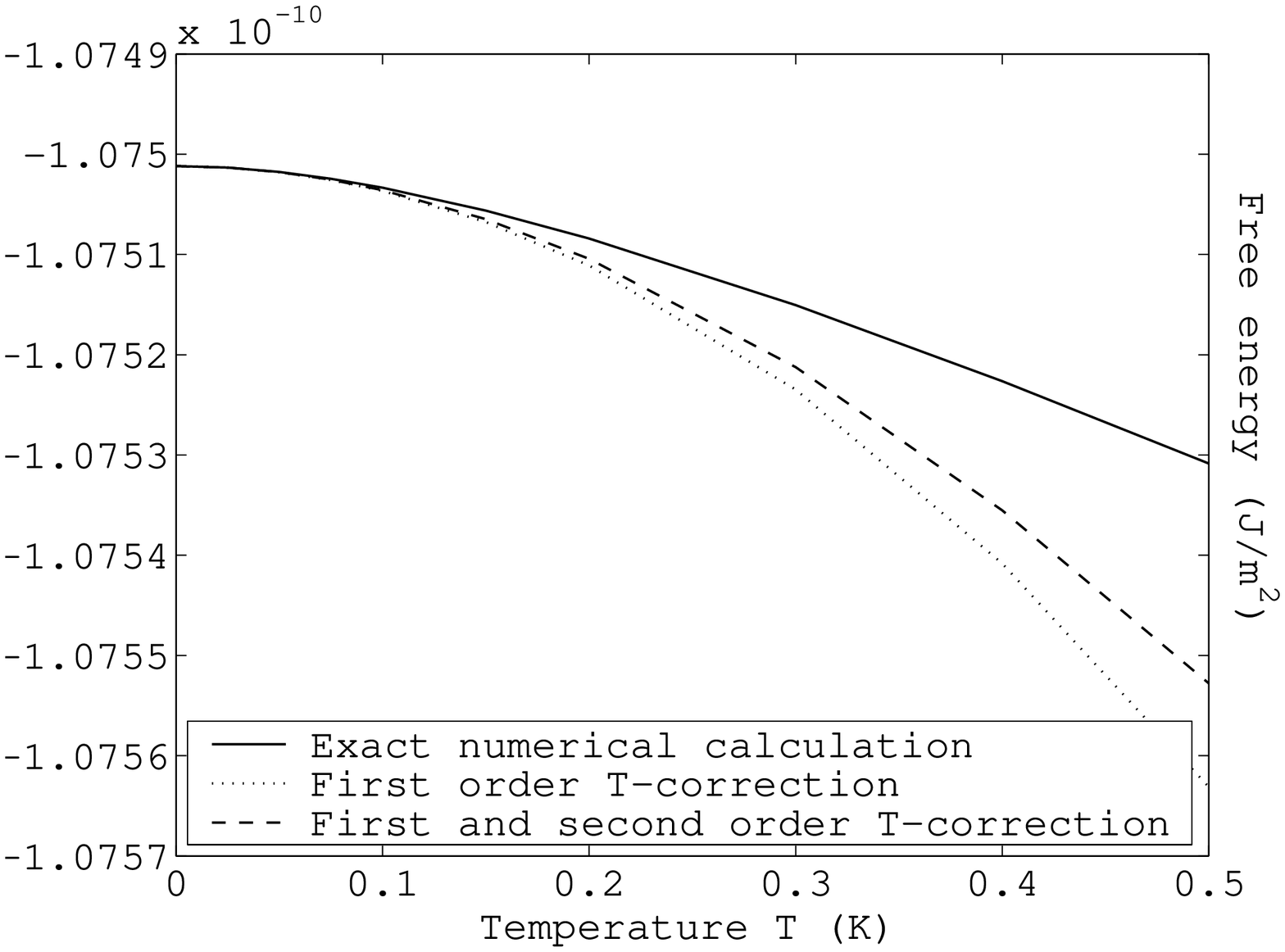}
    \includegraphics[width=3in]{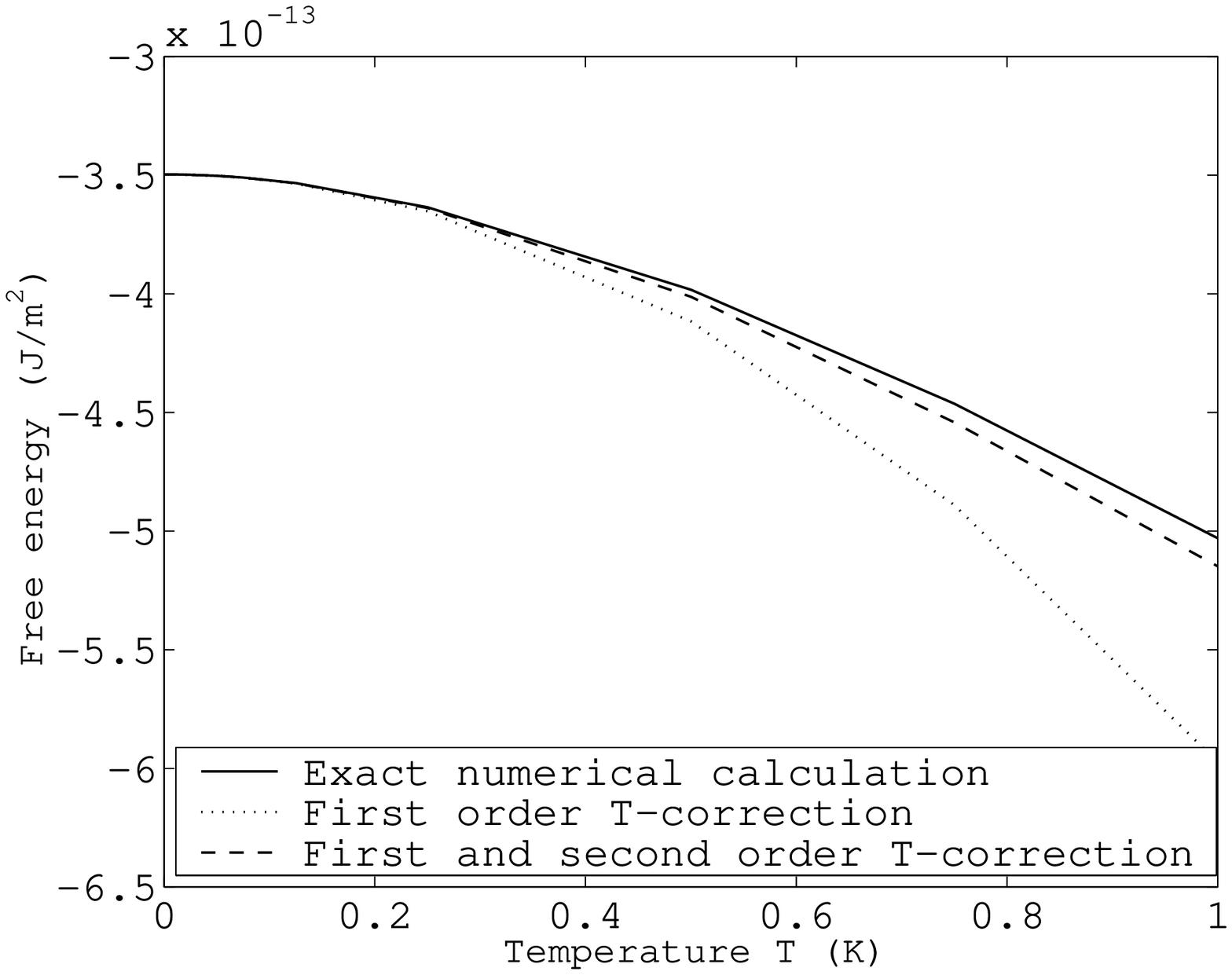}
    \caption{$F^\text{TM}$ and its approximation with (a) $\ez=11.67$ and (b) $\ez=1$. 
    Correction curves, calculated from Eq.~(\ref{dFTM}), are shifted to match 
    the numerical calculations at $T=0$ in each graph.}\label{fig}
  \end{center}
\end{figure}

The numerical investigation in Fig.~\ref{fig}a employs Eq.~(\ref{ep}) with 
$\siSIe=10^{12}$s$^{-1}$, $\ez=11.67$ and $\omega_0=8\cdot10^{15}$s$^{-1}$ as 
appropriate for Si \cite{chen07}. While the analytical expression fits well for $T< 0.1$K, 
corrections $\sim T^4$ become important beyond this point. The two leading orders in temperature 
corrections were shown to be independent of $\ez$ to leading order in 
$\alpha$, but the $T^4$ correction 
(not calculated analytically herein) depends heavily on this value. A qualitative measure of this 
effect is given in Fig.~\ref{fig}b where we have used $\ez=1$, cetera paribus. 
In all plots the theoretical curve is that given in Eq.~(\ref{dFTM}), ignoring corrections.

It is noteworthy that, as seen from Fig.~\ref{fig}, while the $\zeta^{-1}$ term of 
Eq.~(\ref{ep}) gives the dominant temperature correction for small $T$, nearly all 
(99.7\% with our data) of the free energy at $T=0$ is due to the $\ez$ term.


\begin{figure}
  \begin{center}
    \includegraphics[width=3in]{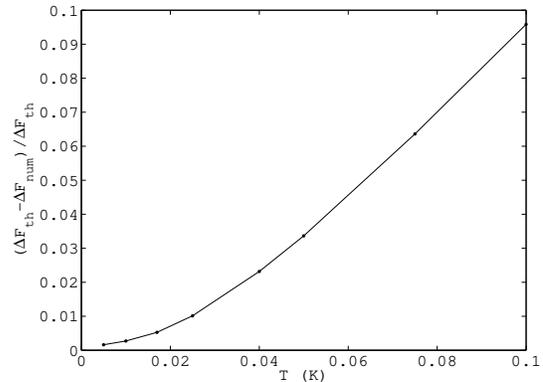}
    \caption{The quantity $R$ defined in Eq.~(\ref{R}) plotted for the TM result 
    (\ref{dFTM}) and numerical calculations.}\label{fig_R}
  \end{center}
\end{figure}

While the fit pictured in Fig.~\ref{fig} is indicative, a much more sensitive 
confirmation of the accuracy of the theoretical results is provided by considering the 
quantity
\be\label{R}
  R = \frac{\Delta F_\text{th}-\Delta F_\text{num}}{\Delta F_\text{th}}
\ee
where $\Delta F_\text{num}$ is the direct numerical calculation and $\Delta F_\text{th}$ 
is the theoretical result to next-to-leading order, in the form (\ref{dFTM}). 
An analysis exactly like this was performed in Ref.~\cite{hoye07}; the reader may refer 
to that paper for further details.

We have found that $\Delta F_\text{th}$ is of the form 
\be
  \Delta F_\text{th} = -CT^2(1-C_1 T)
\ee
and assume $\Delta F_\text{num}$ to be of the form
\be
  \Delta F_\text{num} = -DT^2(1-D_1T+D_2T^2+\dots),
\ee
from which one finds
\begin{align}
  R =& \frac{C-D}{C} -\frac{D}{C}(C_1-D_1)T\notag\\
  &-\frac{D}{C}[D_2+C_1(C_1-D_1)]T^2 + \dots.
\end{align}
In the special case where $C=D$ and $C_1=D_1$, this becomes
\be\label{Rcorrect}
  R = -D_2T^2 + \mathcal{O}(T^3)
\ee
which is zero in the limit $T=0$ and has zero slope in this limit. 
The zero temperature limit of $R$ and its slope thus provide measures of the accuracy 
of the theoretical results: if the $T^2$ coefficient is correct, $R$ should approach 
zero as $T\to 0$, and if the $T^3$ coefficient is correct, the slope of $R(T)$ should 
vanish in this limit as well. We have not taken the corrections (\ref{totalCorr}) 
into account in the plotting of Fig.~\ref{fig_R}.

We have undertaken a numerical study of the behaviour close to zero temperature, 
resulting in the graph of $R$ shown in Fig.~\ref{fig_R}. Due to the vanishing 
denominator of Eq.~(\ref{R}), the analysis is extremely sensitive to numerical errors 
as the zero temperature limit is approached. From the figure it seems clear that the errors in the 
two coefficients are small enough to confirm the correctness of Eq.~(\ref{dFTM}), although 
some caution must be exerted due to the numerical volatility of $R$. Comparing 
Fig.~\ref{fig_R} to Eq.~(\ref{Rcorrect}) it is clear that $D_2<0$ which implies that
the coefficient of the $T^4$ term of the free energy be positive, which conforms with the 
corrections in Fig.~\ref{fig} not accounted for to order $T^3$.

\section{TE mode, residual conductivity}\label{sec_TEcond}

For the TE mode the dominant temperature correction to the free energy comes from the 
last term of Eq.~(\ref{ep}). The permittivity (\ref{ep}), which can be approximated as 
Eq.~(\ref{ep0}),
is similar, but not identical, to that for a Drude metal, considered in 
Refs.~\cite{hoye07, brevik08}. There, instead of Eq.~(\ref{ep0}) the permittivity was 
assumed to be
\be
  \ve_\text{metal} = 1 + \frac{\omega_p^2}{\zeta(\zeta+\nu)} \approx 1 + 
  \frac{\omega_p^2}{\nu\zeta}.
\ee
The principal difference is that the constant term $\ez$ is assumed to be significant here and kept general. 
Since for small $\zeta$ the term $\sim \zeta^{-1}$ dominates the constant term, an 
approximation to the low-temperature behaviour of the dielectric would be expected to be 
found by the same analysis as that of Refs.~\cite{hoye07, brevik08} but with the 
substitution
\be
  \frac{\omega_p^2}{\nu} \to \sigep.
\ee
For typical semiconductors, $\sigep$ is smaller than $\omega_p^2/\nu$ for a good metal by 
many orders of magnitude. For this reason, since the free energy at zero temperature is 
of the same order of magnitude for the metals and semiconductors for the same separation, the 
relative temperature corrections for the TE mode are expected to be much smaller than for 
a metal. Thus, there is reason to investigate whether the effects of $\ez > 1$, while 
negligible for a metal, could be important for small $\sigma$. In some dielectric 
materials, as is well known, $\ez$ can exceed unity by as much as two orders of magnitude, 
and a more careful analysis is therefore justified. The procedure is the same as above, 
and an extension of that found in Ref.~\cite{hoye07}, to which the reader may turn for 
further detail.

It was found in Refs.~\cite{hoye07, brevik08} that for $T\to 0$, and $\ez=1$,
\be\label{dFTE}
  \Delta F_\text{TE} = C_2 T^2 - C_\frac{5}{2}T^{5/2}+\dots,
\ee
where
\begin{subequations}
\begin{align}
  C_2 &= \frac{(\sigep)}{48}(2\ln 2-1),\label{C1}\\
  C_\frac{5}{2} &= \frac{\sqrt{2\pi}}{6}\zeta(-3/2)(\sigep)^{3/2}a.\label{c32}
\end{align}
\end{subequations}
Here $\zeta(y)$ is the Riemann zeta function (for this closed form of $C_\frac{5}{2}$, see 
Appendix A of Ref.~\cite{hoye07}).

For the numerical values indicated this gives the SI values
\begin{subequations}
\begin{align}
  C_2 &= 1.6185719\cdot 10^{-19} \frac{\text{J}}{\text {Km}^2} \left(\frac{\siSIe}{10^{12}\mathrm{s}^{-1}}\right)\\
  C_\frac{5}{2} &= 2.5844373\cdot 10^{-22}\frac{\text{J}}{\text {K}^\frac{5}{2}
  \text{m}^2}\left(\frac{a}{1\mu\text{m}}\right)\left(\frac{\siSIe}{10^{12}  \mathrm{s}^{-1}}\right)^\frac{3}{2}.
\end{align}
\end{subequations}
Thus the TE temperature correction is expected to be positive and in the order of 
magnitude of $10^{-19}$J/m$^{2}$ at $T=1$K. 

\begin{figure}[tb]
  \begin{center}
    \includegraphics[width=3in]{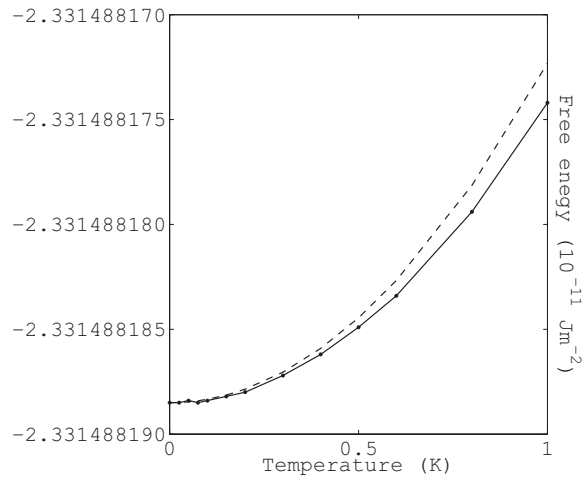}\\
    \caption{
      Temperature dependence of the free energy for TE at 1 micron separation. 
    The solid line is an exact numerical calculation including all terms of Eq.~(\ref{ep})
     with $\ez=11.66$ and $\omega_0=8.0\cdot 10^{15}$s$^{-1}$, the dashed line is the 
       parabolic temperature correction (\ref{dFTE}). The term $\propto T^{5/2}$ is 
       too small to be visible in the graph. 
      }\label{figTE}
  \end{center}
\end{figure}

The numerical calculations shown in Fig.~\ref{figTE}
 were complicated by the fact that the thermal corrections 
are many orders of magnitude smaller than the free energy at zero temperature, making a 
graph of the quantity $R$ similar to Fig.~\ref{fig_R} unfeasible within the 
assumption of $\alpha\ll 1$. We show here, 
however, that assuming $\ez>1$ does not 
change the theoretically predicted thermal correction to 
the free energy to order $T^{3}$, 
and therefore merely refer to 
Ref.~\cite{hoye07} for further numerical support of the 
theoretical result.

\subsection{More general treatment of the TE mode}

Let us treat the TE mode temperature correction to 
the free energy more carefully. 
Starting with the expression (\ref{F}) we perform the substitution
\be
  x = \frac{\kappa}{\zeta\sqrt{\ve(i\zeta)-1}} = \frac{\kappa \mu}{\chi \zeta}
\ee
where we define the recurring quantity
\be
  \chi = \sqrt{\mu + (\ez-1)\mu^2}.
\ee
Then the free energy may be written
\begin{subequations}
\be \label{FTEredo}
  F^\text{TE} = \frac{(\sigep)^3t}{4\pi^2} {\sum_{m =0}^\infty}' g(m)
\ee
with
\be
  g(m) = \chi^2\int_{\mu/\chi}^\infty dxx\ln(1-B e^{-\alpha\chi x}).
\ee
\end{subequations}
The squared reflection coefficient given by (\ref{rTE}) now depends only on $x$:
\be
  B = (x-\sqrt{x^2+1})^4.
\ee

We expand the integrand of $g(m)$: 
\be\label{lnBexp}
  \ln(1-Be^{-\alpha\chi x}) = \ln(1-B) + \frac{\alpha\chi x B}{1-B} + \dots.
\ee
Note that this is as far as we can expand this way, since the next term of the $\alpha$ 
expansion gives a divergent contribution (an alternative method which avoids some 
divergences but is somewhat more cumbersome is the method 
employed in Appendix A of 
Ref.~\cite{hoye07} where the corrections are calculated without the use of the 
Euler-Maclaurin formula.). 

Consider the first terms of the expansion (\ref{lnBexp}) (we dub the terms of $g(m)$ from 
the expansion $g_I(m), g_{II}(m),\dots$):
\be
  g_I(m) = \chi^2\int_{\mu/\chi}^\infty dx\,x\ln[1-(x-\sqrt{x^2+1})^4].
\ee
This integral can be evaluated 
explicitly (a similar integral was evaluated in Ref.~\cite{hoye07} 
where the lower limit was approximated as zero). Perform the substitution 
$x= \sinh u$. Then we may write
\begin{subequations}
\be
  g_I(m) = \frac{\chi^2}{4}\int_{u_0}^\infty du(e^{2u}-e^{-2u})\ln(1-e^{-4u})
\ee
with
\be\label{u0}
  u_0 = \arsinh \frac{\mu}{\chi} = \frac1{2}\ln\left( \frac{\sqrt{\ez\mu+1}+\sqrt{\mu}}
  {\sqrt{\ez\mu+1}-\sqrt{\mu}} \right)
\ee\end{subequations}
With the substitution $y=e^{-2u}$,
\begin{subequations}
\begin{align}
  g_I(m) =& \frac{\chi^2}{8}\int^{y_0}_{0} dy (y^{-2}-1)\ln(1-y^2),
\end{align}
where
\begin{align}
  y_0 =& e^{-2u_0} = \frac{\sqrt{\ez\mu+1}
-\sqrt{\mu}}{\sqrt{\ez\mu+1}+\sqrt{\mu}} \notag\\
  =& 1-2\sqrt{\mu}+2\mu+(\ez-2)\mu^{3/2}+\dots.
\end{align}
\end{subequations}
The integral is straightforward to evaluate and the result is
\begin{align}
  g_I(m) =& -\frac{\chi^2}{8}\left[\left(\frac1{y_0}+y_0\right)\ln(1-y_0^2)\right.\notag\\
  &\left.-2y_0+2\ln\frac{1+y_0}{1-y_0} \right].
\end{align}

We expand this in powers of $\mu$ and find that the terms $\propto \mu^{3/2}$ cancel, consistent with the small-$x$ dependence of the integrand of $g_I(m)$. 
We are left with
\begin{align}
  g_I(m) =&-\frac{\mu}{4}(2\ln 2-1)-\frac{\mu^2}{4}[\ln 4\mu +\ez(2\ln 2-1)] \notag\\
  &+\frac{2}{3}\mu^{5/2} +\order(\mu^3).\label{gIexp}
\end{align}
Comparing with Eq.~(\ref{ggeneral}) we see
\be
  c_1 = -\frac{t}{4}(2\ln 2-1)\text{ and } c_{2l}=-\frac{t^2}{4},
\ee
while the dependence on $\ez$ only enters in the $c_2$ term $\sim m^2$ which does not 
contribute to the Euler-Maclaurin formula. The temperature correction to first order in 
$\alpha$ is thus
\be
  \Delta F_I^\text{TE} \approx \frac{(\sigep)^3t^2}{4\pi^2}\left[\frac{2\ln 2-1}{48}-
  \frac{\Psi t}{4}\right].
\ee
We see that the leading term conforms with Eq.~(\ref{dFTE}) when Eq.~(\ref{C1}) is inserted. 
The first term beyond those calculated is proportional to $T^{7/2}$ according to (\ref{gIexp}). In SI units with (\ref{anPsi}):
\be \label{dFTEredo}
  \Delta F_I^\text{TE} \approx \frac{\siSI(\kB T)^2}{48\varepsilon_0\hbar c^2}(2\ln 2-1)
  -\frac{\zeta(3) (\kB T)^3}{8\pi\hbar^2 c^2}.
\ee

\begin{figure}[tb]
  \begin{center}
    \includegraphics[width=3in]{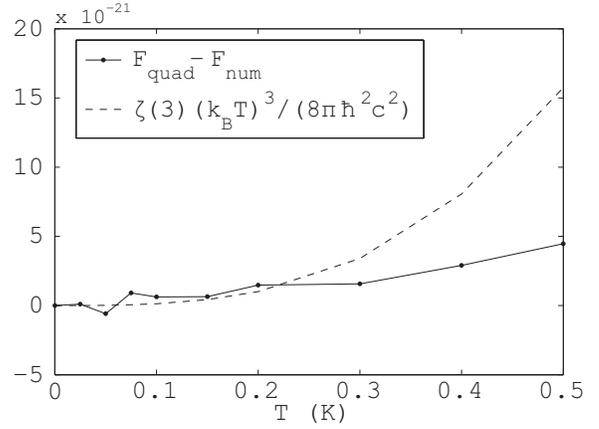}\\
    \caption{
      The difference between the numerically calculated free energy and the 
      quadratic $T$ term of Eq.~(\ref{dFTEredo}) (equal to the difference 
      between the graphs in figure \ref{figTE}) plotted against the absolute 
      value of the $T^3$ term of (\ref{dFTEredo}).
      }\label{fig_cubeTE}
  \end{center}
\end{figure}

The $T^3$ term of (\ref{dFTEredo}) has the same form as that found for 
ideal metals in the limit $aT\ll 1$ \cite{hoye03, milton01}. A similar 
term is present in Eq. (\ref{totalCorr}). [Note that the $T^3$ correction 
for the TM mode is not fully accounted for therein]. A numerical comparison 
of this term with the difference between the graphs in figure \ref{figTE} is 
shown in figure \ref{fig_cubeTE}. It shows that the $T^3$ coefficient in 
(\ref{dFTEredo}) is the right order of magnitude, but the numerical precision 
is not sufficient to draw definite conclusions about its accuracy at this time.

\subsection{First order correction to expansion (\ref{lnBexp})}

The first order correction term in Eq.~(\ref{lnBexp}) is easily calculated with a 
similar scheme. We have
\begin{align}
  g_{II}(m) =& \alpha \chi^3 \int_{\mu/\chi}^\infty dx\frac{x^2B}{1-B} \notag\\
  =& \frac{\alpha\chi^3}{4}\int_{\mu/\chi}^\infty dx \frac{x(x-\sqrt{x^2+1})^2}
  {\sqrt{x^2+1}}.
\end{align}
The procedure for solving this integral is as before. Substitute $x=\sinh u$ to get 
with a little shuffling
\be
  g_{II}(m) = \frac{\alpha \chi^3}{8} \int_{u_0}^\infty du \,e^{-u}(1-e^{-2u}).
\ee
With the substitution $z=e^{-u}$ this becomes very simple:
\be
  g_{II}(m) = -\frac{\alpha \chi^3}{8}\int_{z_0}^0 dz(1-z^2) =
   \frac{\alpha \chi^3}{8}\left(z_0-\frac{z_0^3}{3}\right).
\ee
with 
\begin{align}
  z_0 &= e^{-u_0} = \left(\frac{\sqrt{\ez\mu+1}-\sqrt{\mu}}{\sqrt{\ez\mu+1}+\sqrt{\mu}} 
  \right)^{1/2}\notag\\
  &= 1-\sqrt{\mu}+\frac{\mu}{2} + \frac1{2}(\ez-1)\mu^{3/2}+\dots.
\end{align}
Thus we find the $\mu$ expansion of $g_{II}(m)$:
\be\label{gIIm}
  g_{II}(m) = \frac{\alpha }{8}\left(\frac{2}{3}\mu^{3/2}-(2-\ez)\mu^{5/2}
+\dots\right).
\ee

Hence, with Eqs.~(\ref{generalCorr}) and (\ref{FTEredo}),
\begin{subequations}
\be
\Delta F_{II}^\TE = \frac{(\sigep)^3\alpha }{48\pi^2}\Phi t^{5/2}+\order(t^{7/2}),
\ee
or in SI units with $\Phi=\zeta(-\frac{3}{2})$:
\be  
  \Delta F_{II}^\TE = \frac{\sqrt{2\pi}\zeta(-\textstyle{\frac{3}{2}}) a(\siSIe)^{3/2}}
  {6\hbar^{3/2}}(\kB T)^{5/2}+\dots.
\ee
\end{subequations}
Comparison with Eq.~(\ref{c32}) shows full agreement with the result for metals ($\ez=1$).

It is worth noting that while the next-to-leading temperature correction is of order 
$T^{5/2}$, the term $\propto T^3$ in Eq.~(\ref{dFTEredo}) dominates it with respect to 
$\alpha$. Thus in the small $\sigma$ limit the $T^{5/2}$ dependency becomes all but 
imperceptible.


\section{Conclusions}

We have worked out the two leading terms of the temperature correction to the
Casimir-Lifshitz free energy at low temperatures between poor conductors obeying a 
Drude-type dispersion relation. We have assumed that the materials have a small residual 
conductivity (compared to the reciprocal of the interplate separation) which is finite 
and constant with respect to temperature and frequency near $T=0$. 

The calculation for the TM mode complements that of Ref.~\cite{geyer05} where the 
temperature correction for free energy between two dielectrics of zero conductivity was 
calculated. Both the TE and TM free energy 
temperature corrections are quadratic to leading order. To the extent of our computations, the TM mode has 
integer powers of $T$ beyond the leading whereas the TE mode has both integer and 
half-integer powers. The temperature anomaly reviewed in Sec.~\ref{sec:2} occurs when the 
transition from finite to zero conductivity happens at exactly $T=0$, and while the
analysis presented here does 
not resolve the anomaly, it is of interest to know the low temperature behaviour of 
the free energy in each of the two cases.

Note furthermore that the effects of the static dielectric permittivity $\ez$ only 
enters to order $T^4$ for the TM mode and order $T^{7/2}$ for the TE mode. The fact 
that the coefficient of the term $T^{7/2}$ appears to depend on $\ez$ is noteworthy 
since only integer powers of $T$ were reported in Ref.~\cite{geyer05}, although seeing 
as we have not calculated the coefficient here it is possible that cancellations occur. 

Our calculations are delicate since they rely on the relative smallness of 
different parameters simultaneously. We have assumed the parameter $t$ 
(essentially temperature $T$ divided by conductivity $\sigma$) small while at 
the same time letting $\sigma$ be small compared to the inverse of the 
separation $a$. This is the reason why the leading order temperature 
corrections in (\ref{dFTM}) appear to diverge as $\sigma$ vanishes. On a 
deeper level these subtleties stem from non-commuting limits in the Lifshitz 
formalism which are the cause of anomalies such as that reviewed in section 
\ref{sec:2}. Another curious property both of the present calculations and 
those of Geyer, 
Klimchitskaya, and Mostepanenko \cite{geyer05} is that the free energy 
corrections of order in $T$ just beyond what we have considered here appear 
to diverge as $\ez\to \infty$, as indicated for example by Eqs. (\ref{Li1rel}) 
and (\ref{gIIm}). This limit would a priori be expected to yield the 
ideal metal limit. Such phenomena should be addressed in future studies in the 
effort to achieve 
full understanding of the low temperature behaviour of the Casmir force and 
free energy. 


The asymptotics of the Lifshitz formula as frequency and temperature approach zero are 
fraught with inherent subtleties both mathematical and physical. While the method 
employed herein is highly useful for its simplicity and transparency, it has limitations 
because the functions involved are not analytic in the limits considered and non-integer powers and logarithms enter. 
Physically we have assumed herein a model which may represent certain physical systems, 
but avoids the temperature behaviour which leads to the anomaly reviewed in 
Sec.~\ref{sec:2}. It also neglects effects which may be of importance, 
such as spatial dispersion, a subject which has been extensively
investigated over the years \cite{barton,sernelius,svetovoy,esquivel}. 
A theoretical effort to attempt to describe the screening effects and 
dielectric response of the vanishing density of free charges in insulators near zero 
temperature involving all important physical effects will likely be required in the future and will 
hopefully provide the resolution of the anomaly for dielectrics.

\begin{acknowledgments}
K.A.M.'s research is supported in part by a grant from the US National
Science Foundation (PHY-0554926) and by a grant from the US Department of
Energy (DE-FG02-04ER41305). S.A.E. thanks the University of Oklahoma
for its hospitality while working on this project. We have benefited from discussions and 
suggestions from Emilio Elizalde, Klaus Kirsten, and Jef Wagner.
\end{acknowledgments}

\appendix

\section{Borel summation}\label{app}

The Borel sum of the divergent series $\sum_{n=0}^\infty a_n$ exists 
(Ref.~\cite{bender99}, Sec.~8.2) if the function
\be\label{BorelPhi}
  \phi(x) = \sum_{n=0}^\infty \frac{a_nx^n}{n!}
\ee
is convergent for sufficiently small $x$ and the integral
\be\label{BorelB}
  \mathcal B(x) = \int_0^\infty dt e^{-t}\phi(xt)
\ee
exists. Then the Borel sum is $\sum_{n=0}^\infty a_n = \mathcal B(1)$. Consider the 
quantity $\Psi$ defined in Eq.~(\ref{Psi}) and consider the term 
\be
  \tilde\Psi = \sum_{n=2}^\infty \frac{B_{2n}(2n-4)!}{(2n)!}=\sum_{n=4}^\infty 
  \frac{B_n(n-4)!}{n!}
\ee
(the latter equality follows from $B_{2n+1}=0,\, n=1,2,\dots$). 
Letting $a_{n-4}=B_n(n-4)!/n!$ we get
\be
  \phi(x) = \sum_{n=0}^\infty \frac{B_{n+4}x^n}{(n+4)!}= 
  \frac1{x^4}\left[\frac{x}{e^x-1}-1+\frac{x}{2}-\frac{x^2}{12}\right]
\ee
where the identity $\sum_{n=0}^\infty B_n x^n/n! = x/(e^x-1)$ was used.
The Borel sum
\be
  \tilde\Psi =\mathcal B(1)= \int_0^\infty \frac{dt\,e^{-t}}{t^4}
  \left[\frac{t}{e^t-1}-1+\frac{t}{2}-\frac{t^2}{12}\right]
\ee
is now possible to evaluate analytically 
(the divergence in the lower limit is illusory 
because the expression in brackets is of order $t^4$) to find the desired value as 
$\Psi = 1/36 - 2\tilde \Psi=\zeta(3)/(4\pi^2)$.
A similar numerical procedure using (\ref{BorelPhi}) and (\ref{BorelB}) will give the value of $\Phi$, which as noted in
the text has a well-known asymptotic expression.

An alternative and equivalent approach which is most often simpler is to 
sum each term of the expansion of $g(m)$ in (\ref{ggeneral}) directly 
(disregarding the zero temperature contribution) and obtain finite values 
of the terms of the temperature expansion by means of zeta regularisation
\cite{elizalde94}. Such a procedure immediately yields 
$\Phi=\zeta(-\textstyle{\frac{3}{2}})$ by the definition of the zeta 
function as the analytic continuation of $\zeta(s)=\sum_{m=1}^\infty m^{-s}$. 
Likewise the value of $\Psi$ can easily be found by comparison with the 
asymptotic series expansion of the derivative of $\zeta(s)$:
\be
 \zeta'(s) = -\sum_{n=1}^\infty n^{-s}\ln n, 
\ee
whereby $\Psi = -\zeta'(-2)=\zeta(3)/4\pi^2$. The same reasoning also yields the coefficient of the $c_1$ term of (\ref{generalCorr}) directly as $\zeta(-1)=-1/12$. We thank Emilio Elizalde for alerting us to this point.

\end{document}